\documentstyle[epsfig,psfig,astrobib2,myaasmacros]{mn2e}
\title[Star formation in the HDF-N]
  {Multi-Object Near-IR \Ha\ Spectroscopy of $z\sim1$ star-forming galaxies
    in the HDF-N}
\author[M.Doherty et al.]
 {Michelle Doherty$^1$\thanks{E-mail: md@ast.cam.ac.uk},
Andrew Bunker$^{1,2}$,
Robert Sharp$^{1,3}$,
Gavin Dalton$^4$,
 \newauthor 
Ian Parry$^1$,
Ian Lewis$^4$,
Emily MacDonald$^4$, Christian Wolf$^4$ \& Hans Hippelein$^5$\\
 $^1$Institute of Astronomy, Madingley Road, Cambridge, CB3~0HA, UK\\
 $^2$School of Physics, University of Exeter, Stocker Road, Exeter, UK\\
 $^3$Anglo-Australian Observatory, Sydney, Australia\\
 $^4$Astrophysics, NAPL, Keble Road, Oxford, OX1~3RH, UK\\
 $^5$MPIA, Heidelberg}

\date{Released 2004? Xxxxx XX}

\pagerange{\pageref{firstpage}--\pageref{lastpage}} \pubyear{2004}

\def\LaTeX{L\kern-.36em\raise.3ex\hbox{a}\kern-.15em
    T\kern-.1667em\lower.7ex\hbox{E}\kern-.125emX}

\def\PaA{\ifmmode \mathrm{Pa}{\alpha}\else Pa$\alpha$\fi}
\def\PaB{\ifmmode \mathrm{Pa}{\beta}\else Pa$\beta$\fi}
\def\PaG{\ifmmode \mathrm{Pa}{\gamma}\else Pa$\gamma$\fi}

\def\Ha{\ifmmode \mathrm{H}{\alpha}\else H$\alpha$\fi}
\def\Msol{\ifmmode \mathrm{M}_{\sun}\else M$_{\sun}$\fi}

\begin{document}
 
\label{firstpage}

\maketitle

\begin{abstract}

%put this in conc. in modified form?  
%  We present the first successful demonstration of multi-object
%  near-infrared spectroscopy of high redshift galaxies. 

  We present preliminary results from a programme to obtain multi-object
  near-infrared spectroscopy of galaxies at redshifts $0.7<z<1.5$.
  We are using
  the instrument CIRPASS (the Cambridge Infra-Red PAnoramic Survey
  Spectrograph), in multi-object mode, to survey \Ha\ in galaxies at
  $z\sim1$. We aim to address the true star formation history of the
  universe at this epoch: potentially the peak period of star
  formation activity.  \Ha\ is the same star formation measure
 used at low redshift, and hence we can trace star formation
  without the systematic uncertainties of using different calibrators
  in different redshift bins, or the extreme dust extinction in the
  rest-UV. CIRPASS has been successfully demonstrated in multi-object
  mode on the AAT and WHT. Here we present preliminary results from
  one of our fields, the Hubble Deep Field North, observed with the
  WHT. With 150 fibres deployed over an unvignetted field of $\sim
  15$arcmin, we have several detections of \Ha\ from star forming
  galaxies at $0.8<z<1.0$ and present spectra of the seven brightest of these. By pre-selecting galaxies with redshifts
  such that H$\alpha$ will appear between the OH sky lines, we can
  detect star formation rates of $5\,h^{-2}_{70}\,M_{\odot}\,{\rm
    yr}^{-1}$ ($5\,\sigma$ in 3-hours,
  $\Omega_M=0.3$, $\Omega_{\Lambda}=0.7$ ). 
It appears that star formation rates inferred from \Ha\ are, on average, a
  factor of more than two higher than those based on the UV continuum alone.

\end{abstract}


%%%%%%%%%%%%%%%%%%%%%%%%%%%%%%%%%%%%%%%%%%%%%%%%%%%%%%%%%%%%%%%

\section{Introduction}

A central problem in observational cosmology is determining at which
epoch the majority of stars formed. This has important implications
for models of galaxy formation and evolution. Despite recent progress
in this area, the star formation history of the universe is still a
topic of intense debate. There is substantial evidence that the star
formation rate was much higher in the recent past, compared with the
current epoch, rising steeply to $z\sim1$ (e.g. Lilly et al. 1996,
Tresse et al.  2002, Hippelein et al. 2003)\nocite{llhc96}\nocite{tmlc02}\nocite{hmm+03}. However, it is
still unclear whether at higher redshifts the star formation density
declines, plateaus or perhaps continues to slowly increase. Most
quantitative attempts to measure the global star formation history
have suffered from having to use different indicators of star
formation in various redshift bins \citep{mfd+96}, redshifted into the
optical. These various indicators not only have uncertain relative
calibration but are also affected differently by dust extinction. To
make a reliable comparison, the same tracer of star
formation must be used at high redshift as that used locally. The \Ha\ emission
line is a good tracer of the {\em instantaneous} star formation rate,
and has been widely used in surveys at low redshift. It is particularly
suitable as it is relatively immune to metallicity effects and is much
less susceptible to extinction by dust than the rest-UV continuum and
Lyman-$\alpha$ (which is also selectively quenched through resonant
scattering). It is also important to use \Ha\ to calibrate secondary
indicators of star formation such as [OII]3727\AA, which are often used at
redshift one. However, tracing \Ha\ to early epochs forces a move to
the near-IR at $z>0.6$. The recent advent of good infra-red
spectrographs on large telescopes has facilitated rapid advancement in
this area, with observations of a few tens of galaxies at $z\sim1$
(e.g. Tresse et al., 2002,\nocite{tmlc02} Glazebrook et al.,  1999\nocite{gbe+99}) and
$z\sim2$ \citep{ess+03}. However, until recently,
near-infrared spectroscopy has been restricted to long-slit work and
building samples using single object spectroscopy is inefficient in
terms of telescope time. The small statistical samples obtained result
in large uncertainties in the global properties of galaxies at $z\sim
1$. There has been successful ``multi-object'' spectroscopy through
slitless surveys in the near-infrared from space (the HST/NICMOS
survey of Yan et al.\ 1999\nocite{ymf+99}), but such an approach has poor sensitivity because of the high background. It is only now that true
multi-object infrared spectroscopy is possible from the ground, using
either a slitmask approach (e.g. IRIS\,2 and FLAMINGOS) or a
fibre-fed spectrograph. In this paper we present the first successful
demonstration of multi-object near-infrared spectroscopy of
high-redshift galaxies.

We have used our fibre-fed CIRPASS spectrograph (Cambridge InfraRed
PAnoramic Survey Spectrograph, Parry et al. 2000) \nocite{pmj+00} to
measure the star formation rates of a sample of galaxies in Hubble
Deep Field North (HDF-N) \citep{wbd+96}. This is the initial stage of a
larger survey to address the true star formation history of the universe at redshifts
$z=0.7-1.5$, through \Ha\ measurements of several hundred galaxies.  CIRPASS can operate with
    a 150 fibre multi-object bundle, with the ability to simultaneously observe 75 object+sky
    targets. It thus offers a huge multiplex advantage, compared with
    surveys utilising single object spectroscopy - of which the largest to
    date is the work of Tresse et al. (2002), which surveys a total of 33
    galaxies. Controlled selection of our targets ensures an accurate
    knowledge of the completeness of our survey. The selection function
    will be discussed in a future paper (Doherty et al., 2004, {\it in prep.}).

% -- comparable to the optical 2df spectrograph but working at $z\sim1$ rather than
%$z\sim0.1$

In this paper, we focus on the \Ha\ detections from our pilot study
of galaxies at $z\sim 1$ in the HDF-N with CIRPASS.
The layout of the paper is as follows: in section 2 we describe the
instrument and our observations and in section 3 we detail the data
reduction techniques employed. In section 4 we present results for a sub-sample of
objects, which constitute our brightest detections of \Ha\ (with
$S/N>5\,\sigma$), and use the results to derive the star
formation rates in these redshift one galaxies, comparing these with
that inferred from the rest-frame UV (2400\AA) continuum. Our conclusions are
presented in section 5.

In this paper we adopt the standard ``concordance''
cosmology of $\Omega_M=0.3$, $\Omega_{\Lambda}=0.7$, and use
$h_{70}=H_0/70\,{\rm km\,s^{-1}\,Mpc^{-1}}$. AB magnitudes \citep{og83}
are used throughout.

%%%%%%%%%%%%%%%%%%%%%%%%%%%%%%%%%%%%%%%%%%%%%%%%%%%%%%%%%%%%%%%%%%%%%%%%%

\section{Observations}

\begin{figure}
\begin{tabular}{c}
\hline
\includegraphics[width=50mm]{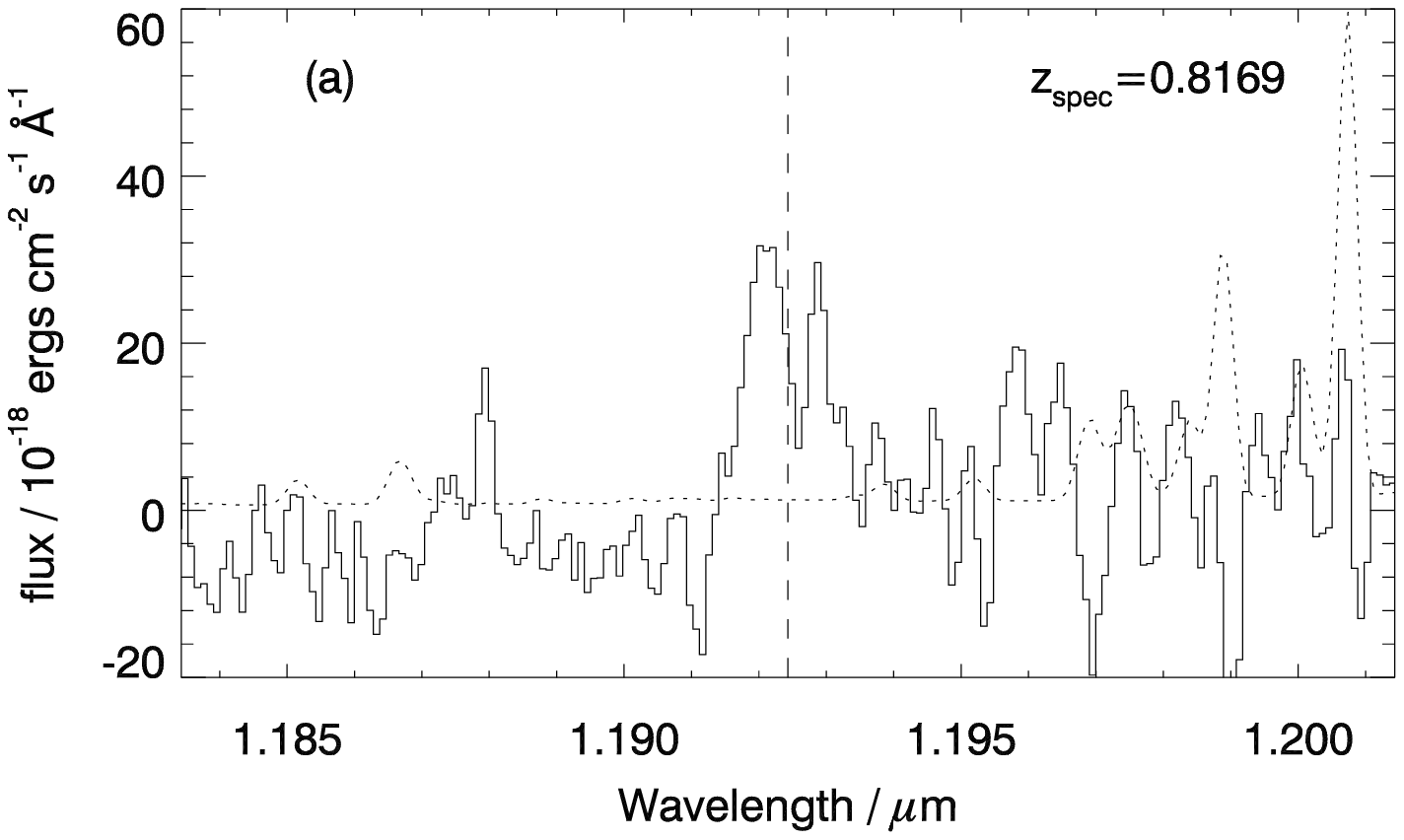} \\
\includegraphics[width=50mm]{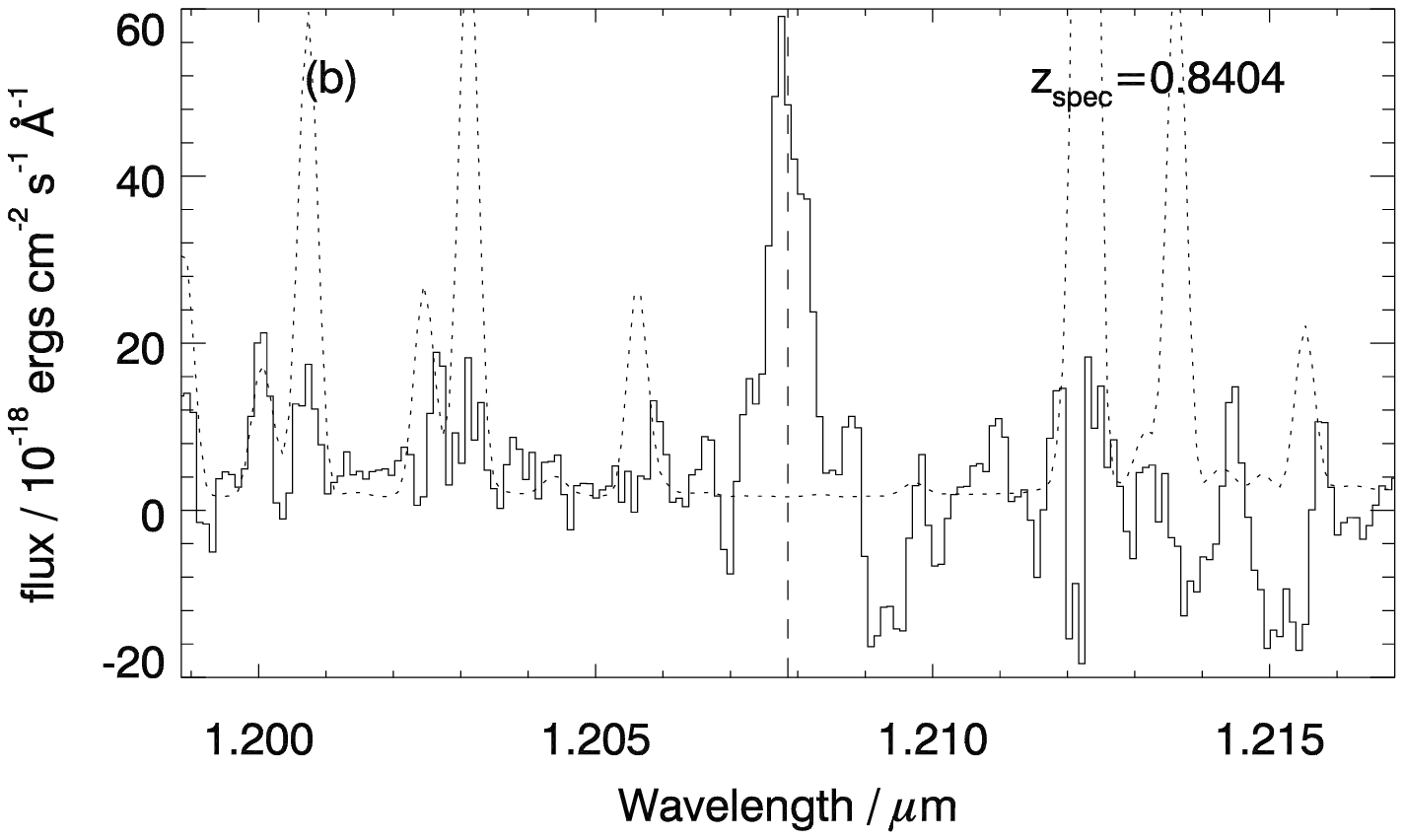}\\ 
\includegraphics[width=50mm]{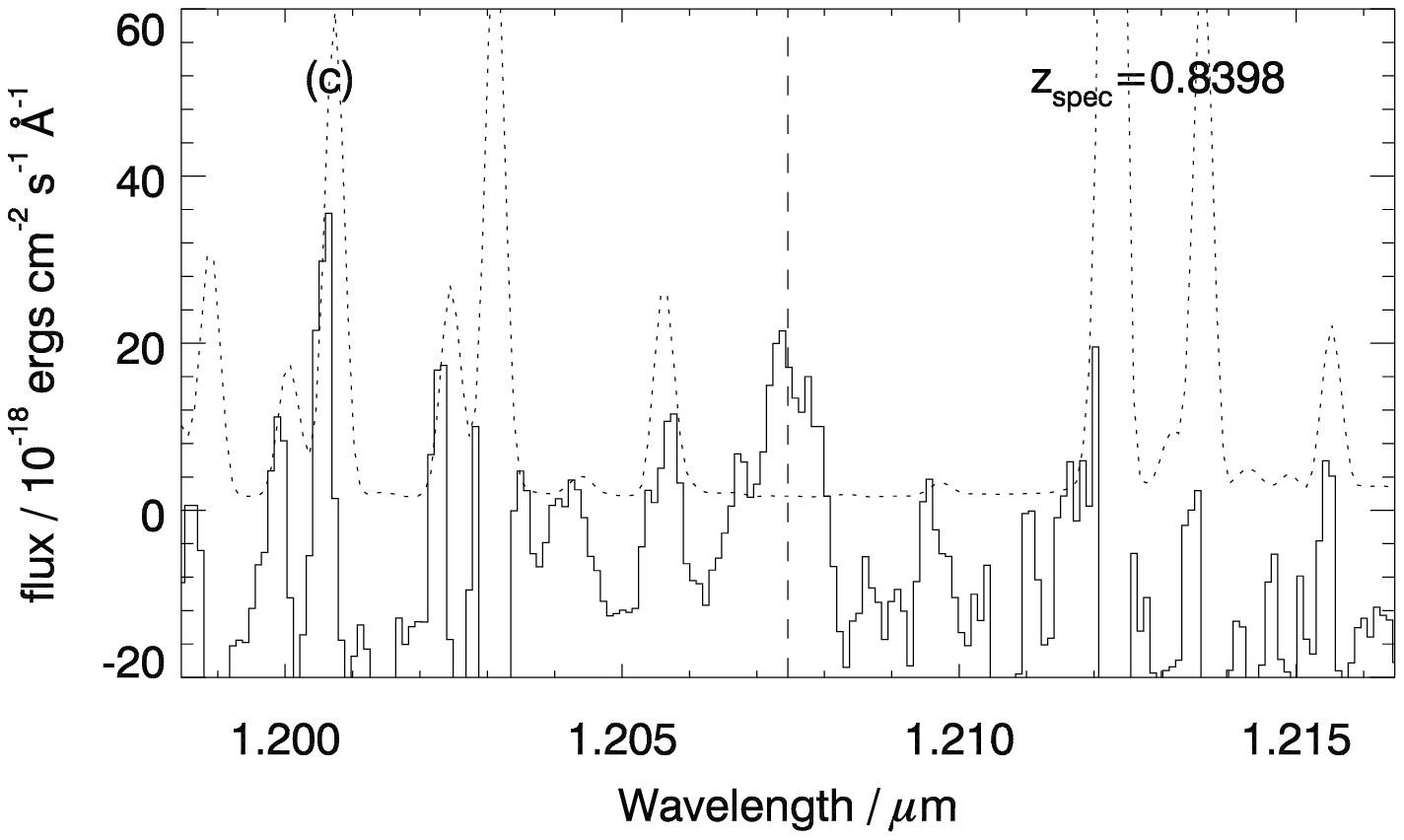} \\
\includegraphics[width=50mm]{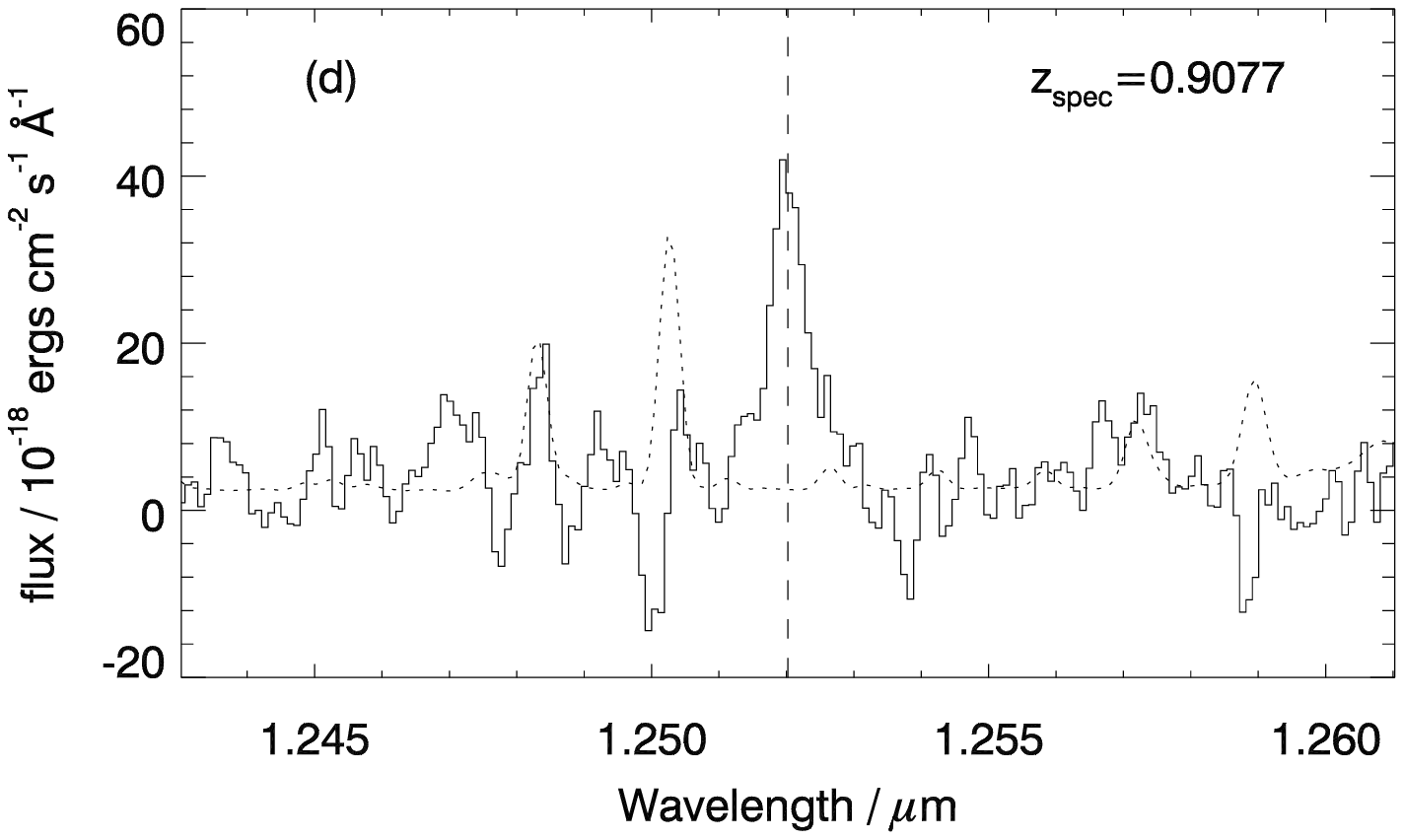} \\  
\includegraphics[width=50mm]{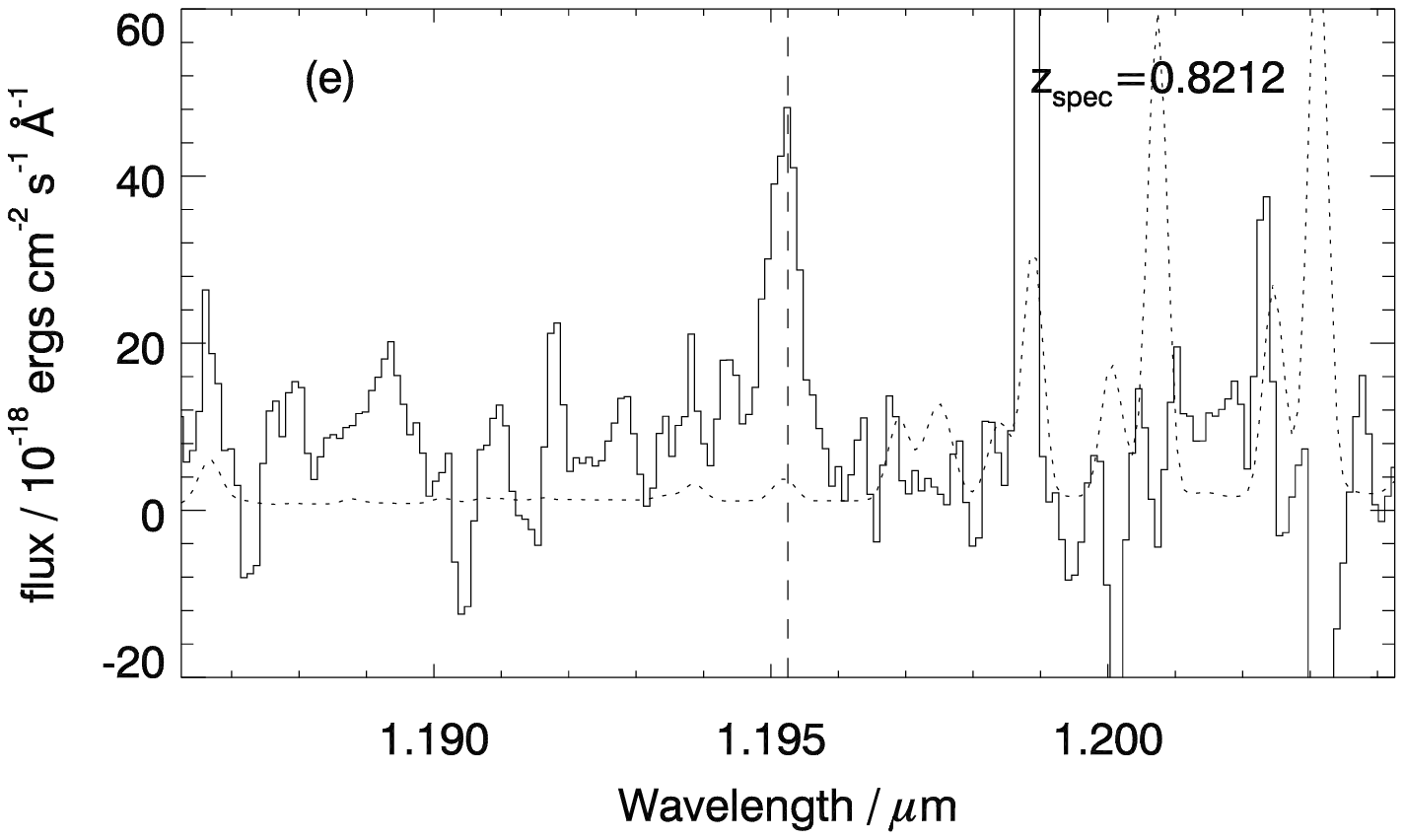}\\
\includegraphics[width=50mm]{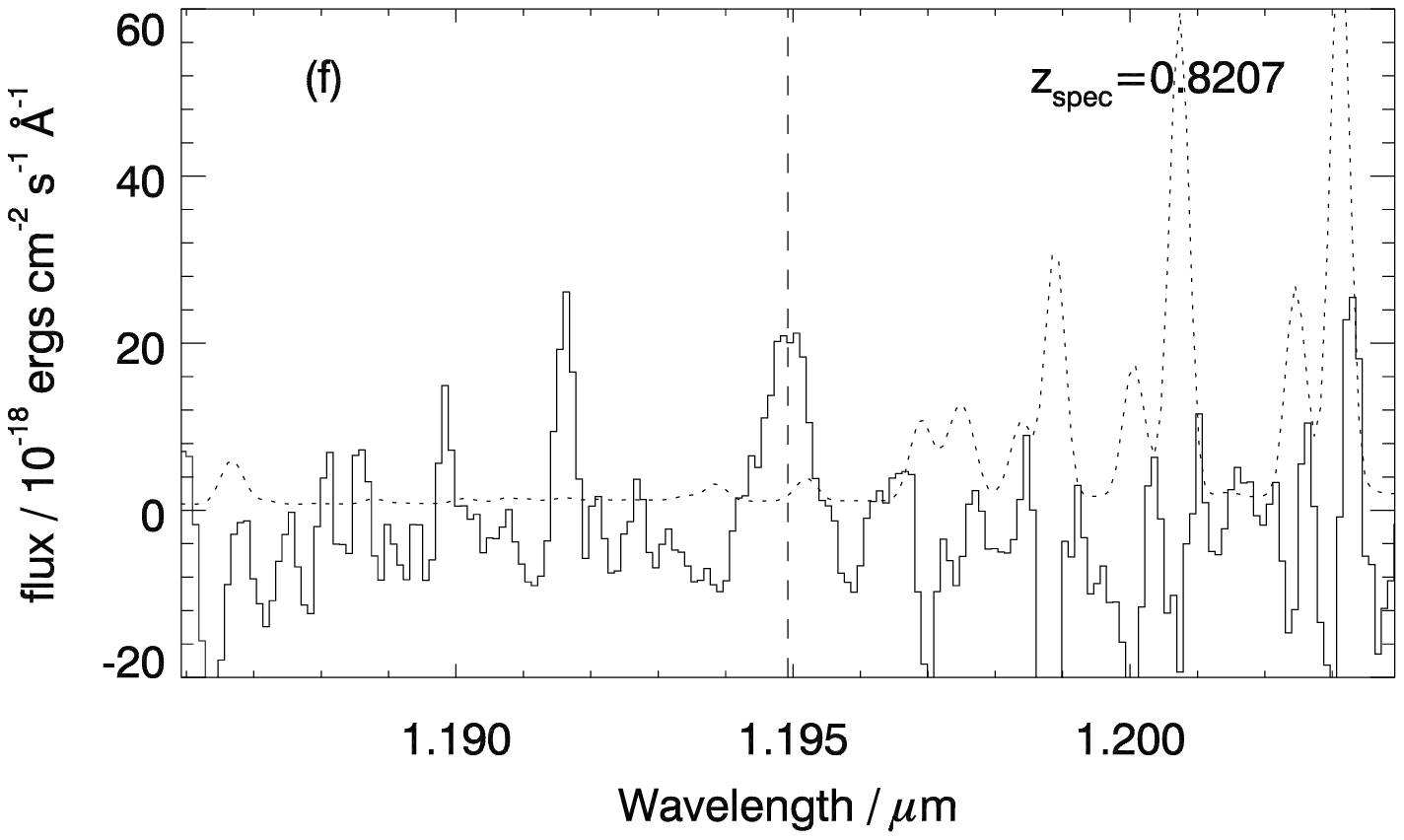}\\ 
\includegraphics[width=50mm]{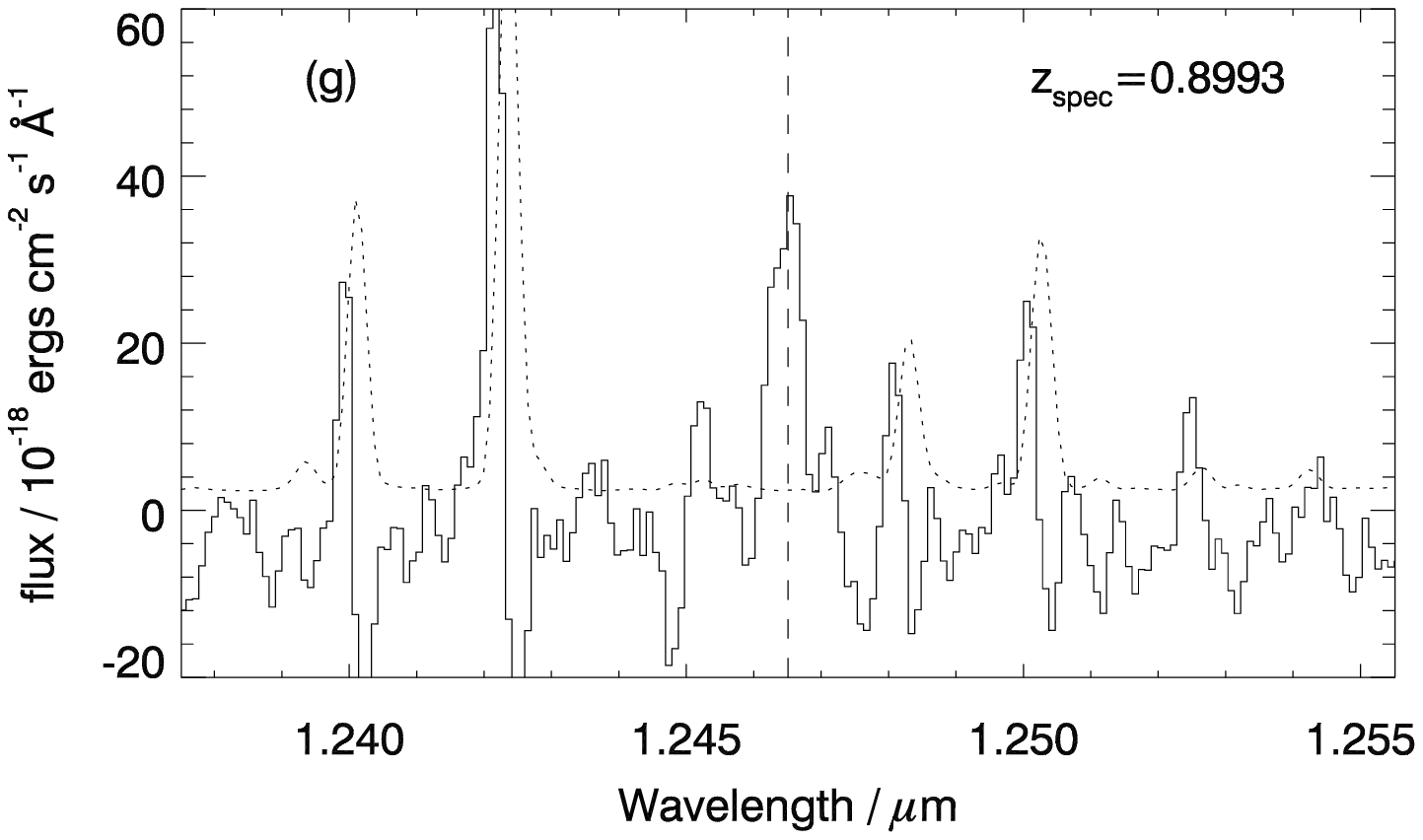}\\

\hline
\end{tabular}
\caption{Spectra for the 7 brightest \Ha\ detections in our sample. The expected position of \Ha\
  at the optical
  spectroscopic redshift is marked with a dashed line. The sky spectrum is
  overlaid in dotted lines. The emission lines all fall well between sky
  lines. }
\label{spectra}
\end{figure}

We performed multi-object spectroscopy of galaxies in the vicinity of
the HDF-N with CIRPASS, a
near-infrared fibre-fed spectrograph operating between 0.9 and 1.67
$\mu$m. The upper cut-off is set by a blocking filter which reduces
the thermal background. These HDF-N observations were undertaken at
the cassegrain focus on the 4.2 m William Herschel Telescope (WHT) in
La Palma. The instrument was used in multi-object mode, with 150
fibres of 1.1\arcsec\ diameter on the sky (comparable to the expected
seeing convolved with typical galaxy profiles). These fibres are deployable over an unvignetted field of
diameter 15$^{\prime}$. We used the FOCAP mechanism developed by the
AAO as an interface to the telescope, with holes drilled in a brass
plate to hold fibres in position. We acquired the field using
six guide bundles, each containing seven closely packed fibres, centred on
bright stars. We thus centroided
the plate to an accuracy better than 0.5\arcsec, that is, less than
half the fibre size.

A Hawaii 2K detector was used, and a grating of 831 lines/mm,
producing a dispersion of 0.95\AA/pixel. The FWHM of each fibre
extends over 2.7 pixels in both the spatial and spectral domain. The
wavelength coverage was 1726\AA, covering most of the J-band, and the
grating was tilted to set a central wavelength of $\lambda_c =
1.25 \mu$m.  The resolving power was $R =
\lambda/\Delta\lambda_{FWHM}\approx5000$.  At this resolution the
background is very dark between the OH sky lines in the J- and
H-bands, with only about 10\% of the array contaminated by skylines.
By targeting galaxies with redshifts which place the \Ha\ emission
lines between the skylines we become very sensitive to line emission
-- indeed, we are limited by the instrument background (although it is
cooled to $-42^{\circ}$C) rather than
the sky. 
%This true if we use other nights obs:
%To further reduce the background in the J-band we put in an additional blocking filter which cuts off off 1.4$\mu m$ was put
%in on the 4th night of the observing run, to reduce thermal and
%background light and increase the sensitivity {\em need a better
%  description of filters here- ask ian}. 
We targeted 65 objects using pairs of fibres offset by 6\arcmin,
nodding the telescope between the A and B positions such that the fibre pairs
alternately received light from the target object and sky. The 2D data
frames are then subtracted directly from each other before extracting the
spectra, so that sky subtracted off has been observed down the same fibre
as the object. Although the sky
background varies a lot on timescales equivalent to our exposure time, we
achieved reasonable first order sky subtraction. Some low throughput fibres were not allocated positions. A total of
12 hours integration time was obtained, over 3 nights, with individual
integrations of 30 or 40 minutes per pointing. 
%(details shown in Table~\ref{obs}). 
The array was read out non-destructively every 10 minutes, with three or
four such loops per pointing to allow
for cosmic ray rejection. There were 10 reads per each 10 minute loop to reduce
the readout noise, ensuring that we achieved background--limited
observations.

The majority of our targets in this field were selected from the
redshift survey of the HDF-N and ``flanking fields'' carried out by
Cohen et al.\ (2000)\nocite{chb+00}, which is $>92$\% complete to $R_{AB}=24$
for objects in the HDF-N proper and $R_{AB}=23$ in the flanking fields.  We
selected all $R_{AB} < 24.5 $ targets within the redshift range $0.73<z<1.0$ which
fell within our 15\arcmin\ unvignetted field of view. A handful of
objects were removed because of the limit on the fibre spacing (a
minimum of 20\arcsec ). We redefined the coordinates of our targets
based on the GOODSv1.0 system (Giavalisco et al., 2003)\nocite{gia03}, for
consistency in our astrometry between the targets and the alignment stars.
The completeness of our sample is discussed in Doherty et al. (2004, in preparation). In this paper
we present observations of the seven sources in HDF-N where \Ha\ was detected at
more than 5$\sigma$. These sources are all in the Cohen et al. (2000)
flanking field, with $R_{AB}<23.0$.
%; the coordinates and redshift information for these objects, from Cohen et
%al. (2000) are listed in Table~\ref{detections}.

%\begin{table}
%\begin{tabular}{ccccc}
%\hline
%\hline
%Date & Exposures (s) & filter  \\
%\hline
%27/12/03  & 8 $\times$ 1800s & clear/cb3 \\
%\hline
%30/12/03 & 4 $\times$ 2400s & cb2/cb3 \\
%30/12/03 & 2 $\times$ 1800s & cb2/cb3 \\
%\hline
%02/01/04 & 2 $\times$ 3000s & cb2/cb3 \\
%02/01/04 & 4 $\times$ 2400s & cb2/cb3 \\
% \hline
%\end{tabular}
%\caption{observations spanning three nights. Not all this detail necessary
%  - what info do we want to put here???}
%\label{obs}
%\end{table}

%\begin{table}
%\begin{tabular}{cccc}
%\hline
%%this is our data for the detections in hdfn:
%id & RA (J2000) & Dec (J2000) & Redshift \\
%\hline
%1  & 12 36 17.536  & +62 14 02.70  & 0.818 \\
%29 & 12 37 00.386  & +62 16 17.27  & 0.913 \\
%37 & 12 37 06.293  & +62 15 18.50  & 0.84  \\
%39 & 12 37 08.381  & +62 15 15.04  & 0.839 \\
%40 & 12 37 08.659  & +62 11 28.52  & 0.907 \\
%47 & 12 37 14.141  & +62 10 44.78  & 0.821 \\
%49 & 12 37 16.631  & +62 10 42.36  & 0.821 \\
%50 & 12 37 16.716  & +62 13 10.54  & 0.898 \\
%\hline
%\end{tabular}
%\caption{Identification, coordinates and spectroscopic redshift, as given
%  in Cohen et al. (2000),
%  for objects with strong \Ha\ emission.}
%\label{detections}
%\end{table}

\section{Data reduction}
Data reduction was performed under IRAF with the CIRPASS
package\footnote{http://www.ast.cam.ac.uk/$\sim$optics/cirpass/docs/install\_cirp\_software.html}.
For each multiple read exposure, the average of each loop was
subtracted from the average of the next loop of non-destructive reads
to form sub-integrations of 10 minutes each at the same pointing.
These sub-integrations were then averaged together using the
\texttt{crreject} algorithm to reject cosmic ray strikes (using limits
of 3$\sigma$ according to gain and readnoise). The frames were then
beam-switched by taking A-B pairs (providing first-order sky
subtraction). We then subtracted of the average residual bias, determined
from the unilluminated region of the array (the bias level of infrared
arrays is known to float with time). The spectrum in each fibre was then optimally
extracted according to the prescription of Johnson et al.
(2002)\nocite{jdp02}, that is, the fibres were profile fitted in triplets (central
fibre and two adjacent fibres) and the pixel values weighted and
summed. This removes the contribution from the two adjacent fibres to the
central fibre. This process produced positive (object-sky) and negative
(sky-object) fibre pairs.

Dome flat fields were also optimally extracted
and used to correct variations in the fibre throughput and the
spectral response of pixels on the array. Wavelength calibration was
performed using an argon arc, fitting 30 lines with a third
order polynomial, resulting in an rms dispersion of 0.1\AA. The
spectra were then rectified to lie on a common wavelength scale.

In order to remove sky residuals which did not fully subtract out in the
beam-switch, a low
order polynomial was fitted and subtracted along each rectified column of
the flat-fielded, extracted spectra. This removed residuals due to temporal
variation of the night sky spectrum. Flux calibration was carried out using
observations of a $J=10.68$ magnitude star from
2MASS which was placed on the same HDF-N plate as our targets. As it was
thus observed simultaneously, this also corrects for
temporal changes in the seeing, which fluctuated a lot over the course
of the observations. As a consistency check we looked at 6 bright stars
placed on the plate in another field from the survey. The fluxes were
consistent to within 3\%, comparing the counts in our CIRPASS spectra with
those predicted from the 2MASS $J-$band magnitudes. 

%{\em talk about when we co-add nights - how do we combine the data? eg
%  inverse variance weighting, compnesate for seeing differences etc...}
%%other detail::
% readnoise, gain, background, throughput measurement, sensitivity
% ??? not here
In an average fibre, between skylines, for a spectrally unresolved
line ($<60$km~s$^{-1}$) we achieve a sensitivity of $7.2\times10^{-17} {\rm erg~s^{-1}~cm^{-2}}$ at 5$\sigma$ in 3 hours. However the emission lines we
detect are typically broader than this ($\sim100-250~{\rm km~s}^{-1}$,
Table~\ref{results}).

\section{Results}

In this paper we analyse a sub-sample of our HDF-N data: those galaxies
with strongly detected (greater than 5$\sigma$) \Ha\ emission. These emission
line spectra are shown in Figure~\ref{spectra}.  The spectrum of the
night sky is overlaid in dotted lines, giving a clear indication of
the positions of skyline residuals, where the noise is significantly higher
due to the enhanced background counts. These residuals do not adversely affect the data as the emission
lines shown here all fall well between skylines. The dashed line shows
the expected position for \Ha, given the optical spectroscopic
redshift. The centroid of \Ha\ is accurate to 2 pixels ($\approx2\AA$), so
the error on the redshift is less than 0.001. There is agreement at this
level with the optical redshifts given in Cohen et al. (2000) and which are listed in Column 5 of
Table~\ref{results}. 

%This is comparable with the agreement found for
%optical/infra-red redshifts in the CFRS (\cite{tmlc02}).

%In all cases the agreement is within 300\,km/s. 

We accurately measured the integrated line fluxes by measuring between zero
power points (Table~\ref{results}), having first subtracted off a continuum fit derived from
regions adjacent to the emission line and unaffected by skylines. We
checked our flux measurements by also fitting gaussian profiles to the
lines, finding consistent values. The gaussian fits also gave the full width at
half maximum (FWHM) for each line (Table~\ref{results}). After subtracting the instrumental
resolution in quadrature from the galaxy line widths, we find
velocity FWHMs for these galaxies in the range $\sim
100-250$\,km\,s$^{-1}$, equivalent to $\sigma_{1D}=40-100\,{\rm
  km\,s^{-1}}$. We note that some of this velocity width may be due
to galactic rotation in the extended galaxies (galaxies J1236175+6214027,
J1237084+6215150 and J1237166+6210424 are the most spatially extended and
also exhibit the broadest lines).

\begin{figure}
\resizebox{80mm}{!}{\includegraphics*[125,80][487,720]{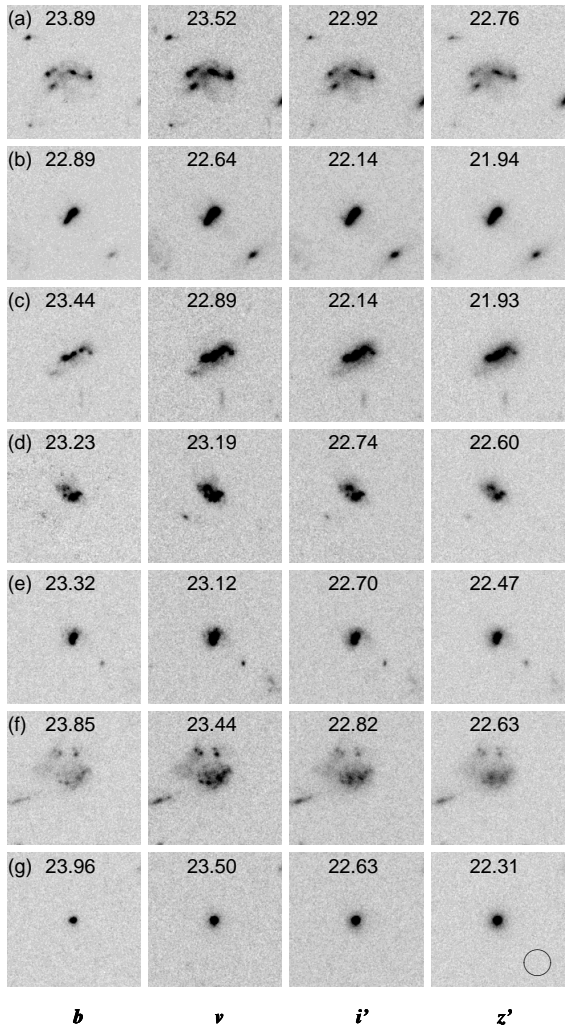}}

\caption{Postage stamp images of the objects showing \Ha\ emission listed
  in Table~\ref{results}, with corresponding identifications. Images are taken from the B,V,i',z' bands in the
  GOODS v1 images. 
%The greyscaling is consistent in each image, with a minimum
%  of -0.006 and maximum of 0.015. 
The image
  size is 5\arcsec to a side. AB magnitudes for 1\arcsec\ apertures (to
  match the fibre size) in each band are shown at the top
 of each image. The size of the fibre on the sky is shown by
  a circle in the bottom
right corner of the stack of images.}
\label{images}
\end{figure}

The \Ha\ luminosity of a galaxy is directly proportional to the
ionising flux from massive stars, which drops off very quickly, about
20 million years after star formation ceases. \Ha\ thus traces the
instantaneous star formation rate, whereas the UV luminosity evolves
in time with the changes in stellar population and continues to rise
even after star formation has ended.  Since the \Ha\ flux is
proportional to the number of OB stars, in order to extrapolate to a
total star formation rate (SFR) it is necessary to assume an initial mass function
(IMF), and the conversion from \Ha\ flux to SFR is quite sensitive to
the IMF assumed (see discussion in Glazebrook et al. 1999). We use Kennicutt's (1998) conversion which assumes a
Salpeter IMF with $0.1\,M_{\odot}<M<100\,M_{\odot}$:
\begin{equation}
{\rm SFR}(\Msol~yr^{-1})=7.9\times10^{-42}~L_{\Ha}~(erg~s^{-1})
\label{sfr_ha}
\end{equation}
The derived SFRs of the galaxies are shown in Table~\ref{results2},
and in Figure~\ref{images} we show $B$, $V$, $i'$ and $z'$ band images
from HST/ACS taken from GOODS v1.0 (Giavalisco et al., 2003)\nocite{gia03}. We
performed photometry on these images with 1\arcsec\ apertures, for
consistency with our infra-red fibre size. In most cases the 1.1\arcsec\ 
fibres enclose most of the B-band light. However, for three of the
sources (galaxies J1236175+6214027,
J1237084+6215150 and J1237166+6210424) some fraction ($<20$\%) is missed.
We reiterate that we have compensated for the effects of seeing by placing
one of the fibres on a bright star in the field (thus observing the star
simultaneously with the galaxies). This allows us to correct the
flux calibration for seeing dependant aperture losses. 

We used the B band magnitudes (4500\AA) to calculate rest frame UV (2400\AA) flux
densities and corresponding star formation rates. These are also shown
in Table~\ref{results2}. For consistency, we also use the conversion
given in Kennicutt (1998)\nocite{ken98}:
\begin{equation}
{\rm SFR}(\Msol~yr^{-1})=1.4\times10^{-28}L_{\nu}(erg~s^{-1}~Hz^{-1})
\label{sfr_uv}
\end{equation}
 assuming the same IMF as used when deriving SFRs from the \Ha\ flux
 (Equation~\ref{sfr_ha}). The UV SFR relation also assumes continuous star
 formation over $\sim10^8$ years. 

\begin{figure}
\includegraphics[width=80mm]{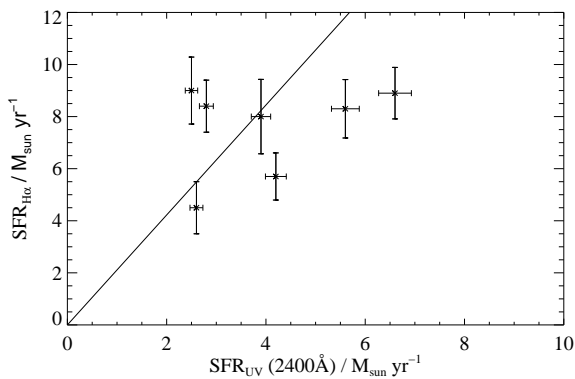}
\caption{Comparison of SFRs obtained from UV continuum flux at 2400\AA\
  versus \Ha\ flux for the individual galaxies. The  SFRs derived from UV luminosity
  are
  consistently underestimated. The solid line has a gradient of 2.11 and represents the line of best fit to the data (using a least
  squares fit). The errors in the SFR(\Ha) values reflect the  errors in
  the flux calibration. We assume $\sim5\%$ error in the UV flux, which is
  dominated by errors in positioning the fibres (i.e. mirrored in the exact
  positioning of the aperture for the $B_{AB}$ magnitude.)}
\label{SFR_plot}
\end{figure}

Figure~\ref{SFR_plot} shows the SFRs for each galaxy calculated using
the UV luminosity density and the \Ha\ flux: those calculated from the UV
luminosity densities are a factor of two lower on average, than those from
\Ha. This is probably due to the differential effect of dust extinction
in the redshift one galaxies between $\lambda_{\rm rest}\approx$ 2400\AA\ and 6563\AA.
This is consistent with results obtained by Glazebrook et al. (1999),
Tresse et al. (2002), and Yan et al. (1999)\nocite{ymf+99} who all
find SFR(\Ha)/SFR(UV) ratios of around 2-3. 

There is no uniform correction that can be applied to the SFR estimated
from UV flux in order to derive the true star formation rate: the amount of extinction varies
in each object due to inherent differences in physical properties of
the galaxies (e.g. \citep{smc+01}). As can be seen in Figure~\ref{SFR_plot}, the statistical errors in
our \Ha\ flux measurements are not great enough to account for the
scatter, which instead is attributable to different dust extinctions in our galaxy
sample. \Ha\ provides a more robust indicator for SFR, giving
an independent estimate that is much less affected by dust
obscuration. The ratio of the total SFR$_{\Ha}$ to SFR$_{UV}$ for this
sub-sample is 2. This fraction is potentially biased high, as there is
a selection effect in this sub-sample due to the fact that we are only including the
sources with strong detections of \Ha : we will address the full
$R$-band magnitude-limited sample in a future paper.

We now estimate the true star formation rate for each galaxy, assuming that
the Kennicutt (1998) relations for \Ha\ and UV SFRs should yield the same
extrinsic SFR in the absence of extinction (i.e. assuming a Salpeter IMF
and continuous star formation). Furthermore, we assume a Calzetti dust
extinction law, appropriate for starburst galaxies \citep{cal97}. For each
galaxy we calculate
the reddening value E(B-V) which gives the observed SFR(\Ha)/SFR(UV) ratio. In
Table~\ref{results2} we list the reddening values derived and use these to
compute the total dust corrected star formation rates. These are typically
50\% higher than those based purely on \Ha, corresponding to an average
$E(B-V)=0.16$. Hence surveys based solely on UV (e.g. 2400\AA) continua (without dust
correction) will underestimate SFRs
at redshift one by factors of $\approx 2-11$.

\begin{table*}
\begin{tabular}{ccccccccccccc}
\hline
%these are our measurements for the detections in hdfn:
id & name& RA & Dec & $\lambda_{\Ha}$ & $z_{\rm opt}$ &$z_{\Ha}$ &fwhm &
   Vel. FWHM & flux  $\times 10^{-16}$ \\
   & &(J2000) & (J2000) &  (\AA) &               &          & (\AA) & km~s$^{-1}$ & ${\rm erg~s^{-1}~cm^{-2}}$\\
\hline
(a)&J1236175+6214027 & 12 36 17.536  & +62 14 02.70 &11924 &0.818 &0.8169 &   *  &  *  & $3.60\pm 5.58$\\
(b)&J1237063+6215185 & 12 37 06.293  & +62 15 18.50 &12078 &0.84  &0.8404 &  5.9 & 127 & $3.31\pm0.38$\\
(c)&J1237084+6215150 & 12 37 08.381  & +62 15 15.04 &12074 &0.839 &0.8398 & 11.1 & 259 & $3.01\pm0.54$\\
(d)&J1237087+6211285 & 12 37 08.659  & +62 11 28.52 &12520 &0.907 &0.9077 &  5.8 & 124 & $2.54\pm0.34$\\
(e)&J1237141+6210448 & 12 37 14.141  & +62 10 44.78 &11952 &0.821 &0.8212 &  4.8 &  97 & $2.27\pm0.36$\\
(f)&J1237166+6210424 & 12 37 16.631  & +62 10 42.36 &11949 &0.821 &0.8207 &  6.6 & 146 & $1.78\pm0.40$\\
(g)&J1237167+6213105 & 12 37 16.716  & +62 13 10.54 &12465 &0.898 &0.8993 &  5.5 & 116 & $2.63\pm0.31$\\
\hline
\end{tabular}
\caption{Measured properties of the \Ha\ emission lines in our $>5\sigma$
 detections in the HDF-N. The optical redshifts are from Cohen et
 al. (2000). The \Ha\ line in J1236175+6214027 is not well fit by a
 gaussian profile, so no FWHM is reported. }
\label{results}
\end{table*}

\begin{table*}
\begin{tabular}{cccccccccccccccc}
\hline
%these are our calculations for the detections in hdfn:
id    &  $R_{AB}$ & SFR$_{\Ha}$     & B$_{AB}$ &flux density $L_{\nu}$            &
      SFR$_{UV}$ & ratio  & E(B-V) & SFR$_{tot}$~\Msol~yr$^{-1}$ \\
      & mag&\Msol~yr$^{-1}$ & mag      & $10^{28} {\rm erg~s^{-1}~Hz^{-1}}$
      &\Msol~yr$^{-1}$&SFR$_{\Ha}$/SFR$_{UV}$ & mag & (dust corrected)\\
\hline
(a) & 21.73&  9.0 & 23.89 & 1.76 & 2.5 & 3.6 & 0.30 & 28.7 \\
(b) & 22.22&  8.9 & 22.89 & 4.70 & 6.6 & 1.3 & 0.07 & 11.9 \\
(c) & 21.62&  8.0 & 23.44 & 2.81 & 3.9 & 2.0 & 0.17 & 15.4 \\
(d) & 22.23&  8.3 & 23.23 & 4.00 & 5.6 & 1.5 & 0.10 & 12.2 \\
(e) & 22.32&  5.7 & 23.32 & 3.01 & 4.2 & 1.4 & 0.09 &  8.1 \\
(f) & 21.59&  4.5 & 23.85 & 1.85 & 2.6 & 1.7 & 0.13 &  7.5 \\
(g) & 22.72&  8.4 & 23.96 & 2.01 & 2.8 & 3.0 & 0.25 & 22.1 \\
\hline
\end{tabular}
\caption{Identification, $R_{AB}$ magnitude (Cohen et al.,~2000), SFR
      derived from \Ha, HST B-band magnitude (1\arcsec\ aperture) and
 corresponding flux density and SFR$_{uv}$ for our $>5\sigma$ detections. }
\label{results2}
\end{table*}

%caption comment:{\em Note: these are not yet properly flux calibrated so y
%  scale in counts/pixel/angstrom?? or some funny units . We should prob. also %mark expected positions of NII. Still need to add this to code These will be r%eproduced when the flux calibration is sorted out}

%%=========================================
%\begin{itemize}
%\item show 1d spectra for each obj.
%\item 2d of best example
%\item co-added 1d spectrum, with annotations showing [NII], [SII] etc
%\item measure Ha fluxes , compare with UV ctm from GOODS B images [aperture
%  correction effects?), OII not available from Cohen et al. 2000 
%\item SEDs from GOODS - are they star-forming? 
%\end{itemize}
%%=========================================

\section{Conclusion}

We have performed multi-object, near-infrared spectroscopy of $z\sim1$
galaxies in the Hubble Deep Field North, using CIRPASS-MOS. These
observations are part of an on-going survey to trace star formation
rates at redshift one. We have presented our brightest detections of
\Ha\ from three hours of observing time, as a demonstration of the
success of this technique. We have shown that we can detect \Ha\ at
sufficient signal to noise to obtain a good estimate of star formation
rates in these galaxies.  By pre-selecting galaxies with H$\alpha$
redshifted between the OH sky lines, we can detect star formation
rates of $5\,h^{-2}_{70}\,M_{\odot}\,{\rm yr}^{-1}$ ($5\,\sigma$ in
3-hours). The star formation rates obtained are higher than those
estimated from UV continuum by a factor of about two, due to dust
obscuration in the UV. This is consistent with previous work in this
area. We have another $\sim60$ sources in this field, at known
redshifts, therefore by stacking the spectra we will be able to obtain
a global star formation rate for $z\sim1$ galaxies in the vicinity of
the HDF-N. Between this and other fields we hope to build up a sample
of several hundred galaxies which will allow us to determine the \Ha\ 
luminosity function at $z\sim1$ and hence address the true star
formation rate at this important epoch. The success of our pilot
multi-object spectroscopic survey bodes well for future instruments,
such as FMOS on Subaru.

%{\em Notes: so far only first night of data shown. we will reduce rest
%soon. main message of this paper is that we have the 1st demo of MOS on a
%large telescope. We have shown the best of our sample, but it is a small
%fraction of the survey. 
%
%We CAN detect Ha at z=1 at sufficient S/N to say
%something useful. We have sufficient spectral resolutin to resolve the
%emission lines so can maybe say something about vel. dispersion/rotation
%curve ... some sort of kinematic info. 
%
%Need to say something about SFRs measured... are these typical? bright? ??
%compare with morphology/Bmags?
%
%These are our brightest detections, yet Glazebrook finds a mean SFR at z=1
%in teh cfrs of 20-60\Msol/yr.... so ours are much less than this. HAve we
%severely mucked up the flux calibration??? but we are reasonably
%self-consistent with our Bmags, finding a simailar ratio to other
%results. how to explain this? cfrs gals brighter than hdf, typically... 
%
%We go to $2\times10^{-17}$ erg/cm/s, (for unresolved line, a bit less for
%Ha), this is $\sim 5$ times fainter than eg. Glazebrook who quotes a
%limiting flux of $\sim10^{-16}$
%
%I'm a bit worried our \Ha\ fluxes aren't high enough - these are the
%brightest in our sample. yet Tresse et al. find mean \Ha\ fluxes of
%$\sim22\times10^{-17}$erg/s/cm2 at redshifts less than 0.3 ????
%
%}

\subsection*{ACKNOWLEDGMENTS}

This paper is based on observations obtained at the William Herschel
Telescope, which is operated by the Isaac Newton Group on behalf of
the UK Particle Physics and Astronomy Research Council. We thank the
WHT staff, in particular Danny Lennon, Kevin Dee, Rene Ruten, Juerg Rey and Carlos Martin,
for their help and support in enabling CIRPASS to be used as a visitor
instrument.  Simon Hodgkin, Elizabeth Stanway and Paul Allen all
assisted in obtaining the observations.  CIRPASS was built by the
instrumentation group of the Institute of Astronomy in Cambridge, UK.
We thank the Raymond and Beverly Sackler Foundation and PPARC
for funding this project. We are indebted to Dave King, Jim Pritchard,
Anthony Horton \& Steve Medlen for contributing their instrument
expertise. We are grateful to Steve Lee and Stuart Ryder at AAO for
assistance in designing the fibre plug plates, and thank the AAO for
the use of the FOCAP fibre unit. The optimal extraction software for
this fibre spectroscopy was written by Rachel Johnson, Rob Sharp and
Andrew Dean.  This research is also partially based on observations
with the NASA/ESA {\sl Hubble Space Telescope}, obtained at the Space
Telescope Science Institute (STScI), which is operated by AURA under
NASA contract NAS 5-26555. These observations are associated with
proposals \#9425\,\&\,9583 (the GOODS public imaging survey). We also
used results from the Caltech Faint Galaxy Redshift Survey in the
HDF-N, and thank Judy Cohen and colleagues for making these catalogues
publically available. MD is grateful for support from the Fellowship
Fund Branch of AFUW Qld Inc., the Isaac Newton Studentship, the
Cambridge Commonwealth Trust and the University of Sydney. ECM acknowledges
the C.K. Marr Educational Trust.

\bibliographystyle{mn2e}
\bibliography{/home/md/LaTeX/BibTeX/myrefs}

\begin{thebibliography}{}

\bibitem[\protect\citeauthoryear{{Calzetti}}{{Calzetti}}{1997}]{cal97}
{Calzetti} D.,  1997, in American Institute of Physics Conference Series {UV
  Opacity in Nearby Galaxies and Application to Distant Galaxies}.
p.~403

\bibitem[\protect\citeauthoryear{{Cohen}, {Hogg}, {Blandford}, {Cowie}, {Hu},
  {Songaila}, {Shopbell} \& {Richberg}}{{Cohen} et~al.}{2000}]{chb+00}
{Cohen} J.~G.,  {Hogg} D.~W.,  {Blandford} R.,  {Cowie} L.~L.,  {Hu} E.,
  {Songaila} A.,  {Shopbell} P.,    {Richberg} K.,  2000, \apj, 538, 29

\bibitem[\protect\citeauthoryear{{Erb}, {Shapley}, {Steidel}, {Pettini},
  {Adelberger}, {Hunt}, {Moorwood} \& {Cuby}}{{Erb} et~al.}{2003}]{ess+03}
{Erb} D.~K.,  {Shapley} A.~E.,  {Steidel} C.~C.,  {Pettini} M.,  {Adelberger}
  K.~L.,  {Hunt} M.~P.,  {Moorwood} A.~F.~M.,    {Cuby} J.,  2003, \apj, 591,
  101

\bibitem[\protect\citeauthoryear{{Giavalisco} \& {GOODS Team}}{{Giavalisco} \&
  {GOODS Team}}{2003}]{gia03}
{Giavalisco} M.,  {GOODS Team} 2003, American Astronomical Society Meeting,
  202,

\bibitem[\protect\citeauthoryear{{Glazebrook}, {Blake}, {Economou}, {Lilly} \&
  {Colless}}{{Glazebrook} et~al.}{1999}]{gbe+99}
{Glazebrook} K.,  {Blake} C.,  {Economou} F.,  {Lilly} S.,    {Colless} M.,
  1999, \mnras, 306, 843

\bibitem[\protect\citeauthoryear{{Hippelein}, {Maier}, {Meisenheimer}, {Wolf},
  {Fried}, {von Kuhlmann}, {K{\" u}mmel}, {Phleps} \& {R{\" o}ser}}{{Hippelein}
  et~al.}{2003}]{hmm+03}
{Hippelein} H.,  {Maier} C.,  {Meisenheimer} K.,  {Wolf} C.,  {Fried} J.~W.,
  {von Kuhlmann} B.,  {K{\" u}mmel} M.,  {Phleps} S.,    {R{\" o}ser} H.-J.,
  2003, \aap, 402, 65

\bibitem[\protect\citeauthoryear{{Johnson}, {Dean} \& {Parry}}{{Johnson}
  et~al.}{2002}]{jdp02}
{Johnson} R.~A.,  {Dean} A.~J.,    {Parry} I.~R.,  2002, in ASP Conf. Ser. 282:
  Galaxies: the Third Dimension {Extracting IFU Spectra}.
p.~531

\bibitem[\protect\citeauthoryear{{Kennicutt}}{{Kennicutt}}{1998}]{ken98}
{Kennicutt} R.~C.,  1998, \araa, 36, 189

\bibitem[\protect\citeauthoryear{{Lilly}, {Le Fevre}, {Hammer} \&
  {Crampton}}{{Lilly} et~al.}{1996}]{llhc96}
{Lilly} S.~J.,  {Le Fevre} O.,  {Hammer} F.,    {Crampton} D.,  1996, \apjl,
  460, L1

\bibitem[\protect\citeauthoryear{{Madau}, {Ferguson}, {Dickinson},
  {Giavalisco}, {Steidel} \& {Fruchter}}{{Madau} et~al.}{1996}]{mfd+96}
{Madau} P.,  {Ferguson} H.~C.,  {Dickinson} M.~E.,  {Giavalisco} M.,  {Steidel}
  C.~C.,    {Fruchter} A.,  1996, \mnras, 283, 1388

\bibitem[\protect\citeauthoryear{{Oke} \& {Gunn}}{{Oke} \& {Gunn}}{1983}]{og83}
{Oke} J.~B.,  {Gunn} J.~E.,  1983, \apj, 266, 713

\bibitem[\protect\citeauthoryear{{Parry}, {Mackay}, {Johnson}, {McMahon},
  {Dean}, {Ramaprakash}, {King}, {Pritchard}, {Medlen}, {Sabbey}, {Ellis} \&
  {Aragon-Salamanca}}{{Parry} et~al.}{2000}]{pmj+00}
{Parry} I.~R.,  {Mackay} C.~D.,  {Johnson} R.~A.,  {McMahon} R.~G.,  {Dean} A.,
   {Ramaprakash} A.~N.,  {King} D.~L.,  {Pritchard} J.~M.,  {Medlen} S.~R.,
  {Sabbey} C.~S.,  {Ellis} R.~S.,    {Aragon-Salamanca} A.,  2000, in Proc.
  SPIE Vol. 4008, p. 1193-1202, Optical and IR Telescope Instrumentation and
  Detectors, Masanori Iye; Alan F. Moorwood; Eds. {CIRPASS: a NIR integral
  field and multi-object spectrograph}.
pp 1193--1202

\bibitem[\protect\citeauthoryear{{Sullivan}, {Mobasher}, {Chan}, {Cram},
  {Ellis}, {Treyer} \& {Hopkins}}{{Sullivan} et~al.}{2001}]{smc+01}
{Sullivan} M.,  {Mobasher} B.,  {Chan} B.,  {Cram} L.,  {Ellis} R.,  {Treyer}
  M.,    {Hopkins} A.,  2001, \apj, 558, 72

\bibitem[\protect\citeauthoryear{{Tresse}, {Maddox}, {Le F{\` e}vre} \&
  {Cuby}}{{Tresse} et~al.}{2002}]{tmlc02}
{Tresse} L.,  {Maddox} S.~J.,  {Le F{\` e}vre} O.,    {Cuby} J.-G.,  2002,
  \mnras, 337, 369

\bibitem[\protect\citeauthoryear{{Williams}, {Blacker}, {Dickinson}, {Dixon},
  {Ferguson}, {Fruchter}, {Giavalisco}, {Gilliland}, {Heyer}, {Katsanis},
  {Levay}, {Lucas}, {McElroy}, {Petro}, {Postman}, {Adorf} \&
  {Hook}}{{Williams} et~al.}{1996}]{wbd+96}
{Williams} R.~E.,  {Blacker} B.,  {Dickinson} M.,  {Dixon} W.~V.~D.,
  {Ferguson} H.~C.,  {Fruchter} A.~S.,  {Giavalisco} M.,  {Gilliland} R.~L.,
  {Heyer} I.,  {Katsanis} R.,  {Levay} Z.,  {Lucas} R.~A.,  {McElroy} D.~B.,
  {Petro} L.,  {Postman} M.,  {Adorf} H.,    {Hook} R.,  1996, \aj, 112, 1335

\bibitem[\protect\citeauthoryear{{Yan}, {McCarthy}, {Freudling}, {Teplitz},
  {Malumuth}, {Weymann} \& {Malkan}}{{Yan} et~al.}{1999}]{ymf+99}
{Yan} L.,  {McCarthy} P.~J.,  {Freudling} W.,  {Teplitz} H.~I.,  {Malumuth}
  E.~M.,  {Weymann} R.~J.,    {Malkan} M.~A.,  1999, \apjl, 519, L47

\end{thebibliography}

\label{lastpage}

\end{document}